\documentclass[letterpaper,twocolumn,english,american,prl,  showpacs]{revtex4}
\usepackage[T1]{fontenc}
\usepackage[latin1]{inputenc}
\usepackage{amsmath}
\usepackage{graphicx}
\usepackage{amssymb}

\makeatletter

\newcommand{\noun}[1]{\textsc{#1}}

\DeclareMathOperator{\Pre}{Pre}
\DeclareMathOperator{\Post}{Post}

\newcommand{\eps}{\varepsilon}

\newcommand{\mysubstack}[2]{\substack{#1\\#2} }
\newcommand{\MyRe}{\mathrm{Re}}

\usepackage{babel}
\makeatother
\begin{document}

\title{Topological Speed Limits to Network Synchronization}

\author{Marc Timme, Fred Wolf, and Theo Geisel}

\address{Max-Planck-Institut für Strömungsforschung, 37073 Göttingen, Germany}

\begin{abstract}
We study collective synchronization of pulse-coupled oscillators interacting
on asymmetric random networks. We demonstrate that random matrix theory
can be used to accurately predict the speed of synchronization in
such networks in dependence on the dynamical and network parameters.
Furthermore, we show that the speed of synchronization is limited
by the network connectivity and stays finite, even if the coupling
strength becomes infinite. In addition, our results indicate that
synchrony is robust under structural perturbations of the network
dynamics. 
\end{abstract}


\maketitle
Complex networks have attracted an enormous research interest in the
recent past \cite{Albert_et_al}. Most studies have focussed on the
structures of a variety of systems such as the world wide web, email
networks, genetic networks, and biological neural networks \cite{Albert_et_al,Strogatz}.
An equally important task is to understand the \textit{collective
dynamics} on such networks. In particular, the question arises: how
is the dynamics on a complex network influenced by its structure \cite{Strogatz}?

Synchronization appears to be one of the simplest kinds of collective
dynamics among coupled dynamical systems \cite{Pikovsky,Nadis}. It
was found to be ubiquitous in artificial as well as natural networks
as different as Josephson junction arrays \cite{Wiesenfeld} and biological
neural networks \cite{syn}. To understand the dynamics of such networks,
theoretical studies have emphasized systems consisting of simple units
such as phase- and pulse-coupled limit-cycle oscillators \cite{Winfree_et_al,Abbott_et_al}.
Yet, although most real-world networks often possess a complex connectivity
structure, most studies of synchronization of coupled oscillators
are either restricted to networks of globally coupled units and simple
regular networks, or work in some mean field limit \cite{Winfree_et_al,Abbott_et_al}.
Although exact results on synchronization in networks with a general
structure have been obtained recently \cite{Baharona,Earl,TimmePRL},
it is still not well understood how the structure of a complex network
affects dynamical features of synchronization.

In this Letter, we study the collective synchronization of pulse-coupled
oscillators interacting on asymmetric random networks. We find that
the speed of synchronization is restricted by the network connectivity
and stays finite, even if the coupling strength becomes infinite.
No such speed limit exists in large networks of globally coupled units.
More generally, we show that the theory of random matrices can be
used to successfully predict the speed of synchronization as a function
of dynamical and network parameters. In addition, our results indicate
that synchrony occurs robustly, i.e. persists under structural perturbation
of the network dynamics. 

We consider asymmetric random networks of $N$ oscillators which interact
by sending and receiving pulses \cite{Mirollo_et_al}. The sets $\Pre(i)$
of presynaptic oscillators that send pulses to oscillator $i$ specify
the structure of such a network. For each oscillator $i$, the $k_{i}:=|\Pre(i)|$
presynaptic oscillators are drawn from the uniform distribution among
all other oscillators $\{1,\ldots,N\}\backslash\{ i\}$. 

A phase-like variable $\phi_{i}(t)$ specifies the state of each oscillator
$i$ at time $t$. In the absence of interactions, the dynamics of
unit $i$ is given by $d\phi_{i}/dt=1$. When oscillator $i$ reaches
a threshold, $\phi_{i}(t)=1$, its phase is reset to zero, $\phi_{i}(t^{+})=0$,
and the oscillator is said to 'fire'. A pulse is sent to all postsynaptic
oscillators $j\in\Post(i)$ which receive this signal after a delay
time $\tau$. The incoming signal induces a phase jump $\phi_{j}((t+\tau)^{+}):=U^{-1}(U(\phi_{j}(t+\tau))+\eps_{ji})$
which depends on the instantaneous phase $\phi_{j}(t+\tau)$ of the
postsynaptic oscillator and the coupling strength $\eps_{ji}$ which
we take to be inhibitory (phase-retarding), $\eps_{ji}\leq0$. The
phase dependence is determined by a twice continuously differentiable
'potential' function $U(\phi)$ that is assumed to be strictly increasing,
$U'(\phi)>0$, concave (down), $U''(\phi)<0$, and normalized such
that $U(0)=0$, $U(1)=1$ (cf.\ \cite{Mirollo_et_al}). We focus
on the specific form $U(\phi)=U_{\mathrm{IF}}(\phi)=I(1-e^{-T_{\mathrm{IF}}\phi})$
that represents the integrate-and-fire oscillator defined by the differential
equation $\dot{V}=I-V$ (and a threshold at $V=1$). Here $I>1$ is
an external input and $T_{\mathrm{IF}}=\log(I/(I-1))$ the intrinsic
period of an oscillator. Other forms of $U(\phi)$ give qualitatively
similar results. In such a network the synchronous state, $\phi_{i}(t)=\phi_{0}(t)$
for all $i$, exists if the coupling strengths are normalized such
that $\sum_{j\in\Pre(i)}\eps_{ij}=\eps$. Its period is given by $T=\tau+1-U^{-1}(U(\tau)+\eps)$. 

\begin{figure}[htbp]
\includegraphics[%
  width=0.80\columnwidth]{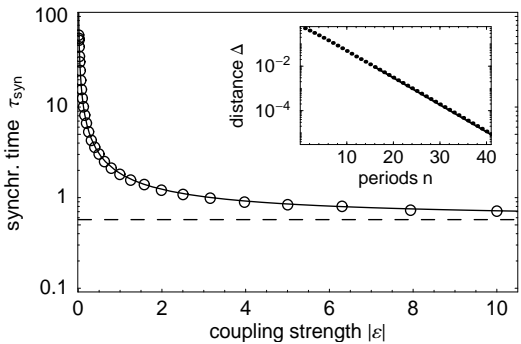}

\caption{Asymptotic synchronization time in a random network ($N=1024$, $k_{i}\equiv k=32$,
$I=1.1$, $\tau=0.05$, $\eps_{ij}=\eps/k$ for $j\in\Post(i)$) .
The inset shows the distance $\Delta$ of a perturbation $\boldsymbol{\delta}$
from the synchronous state versus the number of periods $n$ ($\eps=-0.4$).
Its slope yields the synchronization time $\tau_{\textrm{syn}}$ shown
in the main panel as a function of coupling strength $|\eps|$. Simulation
data ($\bigcirc$), theoretical prediction (------) derived in this
Letter, its infinite coupling strength asymptote (--~--~--). \label{Fig:tausyn}}
\end{figure}

In numerical simulations of the network dynamics, we find that the
synchronous state is always stable, independent of the parameters
(cf.~\cite{TimmePRL}). A sufficiently small perturbation $\boldsymbol{\delta}(0)\equiv\boldsymbol{\delta}=(\delta_{1},\ldots,\delta_{N})^{\mathsf{{T}}}$
of the phases, defined by $\phi_{i}(0)=\phi_{0}(0)+\delta_{i}$ asymptotically
decays exponentially with time. Thus, denoting $\boldsymbol{\delta}'(t):=\boldsymbol{\delta}(t)-\lim_{s\rightarrow\infty}\boldsymbol{\delta}(s)$,
the distance $\Delta(n):=\max_{i}|\delta'_{i}(nT)|/\max_{i}|\delta'_{i}(0)|$
from the synchronous state behaves as\begin{equation}
\Delta(n)\sim\exp(-n/\tau_{\textrm{syn}})\label{eq:delta_tau}\end{equation}
defining a synchronization time $\tau_{\textrm{syn}}$ in units of
the collective period $T$. The speed of synchronization $\tau_{\textrm{syn}}^{-1}$
strongly depends on the parameters. For instance, as might be expected,
synchronization is faster for stronger coupling. Surprisingly, however,
we find that synchronization cannot be faster than an upper bound
even if the coupling strength becomes arbitrarily large (cf.\  Fig.~\ref{Fig:tausyn}). 

 To understand how the speed of synchronization depends on the dynamical
and network parameters, we analyze the linear stability of the synchronous
state. Following \cite{TimmePRL} we obtain a nonlinear stroboscopic
map $\boldsymbol\delta(T)=F(\boldsymbol\delta)$ for the perturbations,
the linearization of which reads\begin{equation}
\boldsymbol\delta(T)\doteq A\boldsymbol\delta\label{eq:matrixequation}\end{equation}
where the elements of the stability matrix $A$ are given by $A_{ij}=p_{i,n}-p_{i,n-1}$
if $j=j_{n}\in\Pre(i)$, $A_{ii}=p_{i,0}$, and $A_{ij}=0$ otherwise.
Here $j_{n}$ identifies the $n^{\mathrm{th}}$ pulse received during
this cycle by oscillator $i$ and $p_{i,n}:=U'(U^{-1}(U(\tau)+\sum_{m=1}^{n}\eps_{ij_{m}}))/U'(U^{-1}(U(\tau)+\eps))$
for $n\in\{0,1,\ldots,k_{i}\}$. For $U(\phi)=U_{\textrm{IF}}(\phi)$
and coupling strengths $\eps_{ij}=\eps/k_{i}$ for $j\in\Pre(i)$
the matrix elements reduce to \cite{uniqueness}

\begin{equation}
A_{ij}=\left\{ \begin{array}{ll}
\frac{1-A_{0}}{k_{i}} & \mbox{if}\, j\in\Pre(i)\\
A_{0} & \mbox{if}\, j=i\\
0 & \mbox{if}\, j\notin\Pre(i)\cup\{ i\}\end{array}\right.\label{eq:IFmatrixelements}\end{equation}
where \begin{equation}
A_{0}=\frac{Ie^{-\tau T_{\textrm{IF}}}}{Ie^{-\tau T_{\textrm{IF}}}-\eps}>0.\label{eq:A0}\end{equation}
 Obviously, $A$ constitutes a row-stochastic matrix, i.e.\  $\sum_{j}A_{ij}=1$
for all $i$. Thus $A$ has one trivial eigenvalue $\lambda_{1}=1$
associated with the eigenvector $\boldsymbol v_{1}=(1,1,\ldots,1)^{\textsf{T}}$
representing a uniform phase-shift and thus reflecting time-translation
invariance. Furthermore, the Gershgorin theorem \cite{Stoer} implies
that all eigenvalues are located inside a disk of radius $r_{\textrm{G}}=1-A_{0}$
centered at $A_{0}$, such that in particular $\left|\lambda_{i}\right|\leq1$
and the synchronous state is (at least neutrally) stable. For simplicity,
we consider networks of homogenous random connectivity, $k_{i}=k$
for all $i$, in the following.

\begin{figure}[htbp]
\begin{center}\includegraphics[%
  width=0.80\columnwidth]{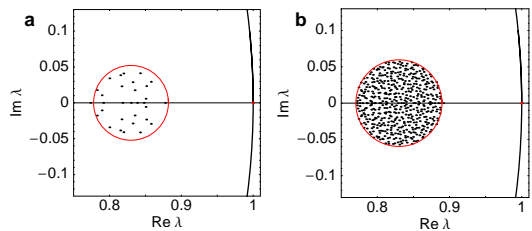}\end{center}

\caption{Distribution of eigenvalues $\lambda_{i}$ of two stability matrices
$A$ in the complex plane ($I=1.1$, $\eps=-0.2$, $\tau=0.05$ $\Rightarrow$$A_{0}\approx0.83$;
$k=8$) for networks of (a) $N=32$, (b) $N=512$ oscillators. For
large networks, the non-trivial eigenvalues seem to be distributed
uniformly on a disk in the complex plane. The prediction from random
matrix theory (Eq.~(\ref{eq:rRMT})) is indicated by a circle. The
arc through the trivial eigenvalue $\lambda_{1}=1$ is a sector of
the unit circle. \label{Fig:EWSfreek8N}}
\end{figure}

We numerically determined the eigenvalues of different stability matrices
$A$ for various network sizes $N\in\{2^{6},\ldots,2^{14}\}$, in-degrees
$k\in\{2,\ldots2^{8}\}$, and dynamical parameters $\eps$, $\tau$,
and $I$ such that $0<A_{0}<1$. In general, we find that for sufficiently
large $k$ and $N$ the non-trivial eigenvalues resemble a disk in
the complex plane that is centered at about $A_{0}$ and has a radius
$r$ that is smaller than the upper bound given by the Gershgorin
theorem, $r<1-A_{0}.$ Examples are shown in Fig.~\ref{Fig:EWSfreek8N}.

\selectlanguage{english}

\selectlanguage{american}
This eigenvalue distribution is reminiscent of the {}``circle law''
of random matrix theory \cite{Girko_et_al}: Gaussian asymmetric random
matrices, having a distribution of matrix elements \begin{equation}
p_{\textrm{Gauss}}(J_{ij})=N^{\frac{1}{2}}(2\pi r^{2})^{-\frac{1}{2}}\exp\left(-\frac{NJ_{ij}^{2}}{2r^{2}}\right)\label{eq:prob_Gauss}\end{equation}
with independent $J_{ij}$ and $J_{ji}$, also exhibit eigenvalue
distributions \begin{equation}
\rho_{\textrm{Gauss}}^{\textrm{a}}(\lambda)=\left\{ \begin{array}{ll}
(\pi r^{2})^{-1} & \textrm{if }|\lambda|\leq r\\
0 & \textrm{otherwise}\end{array}\right.\label{eq:uniform_EW_distribution}\end{equation}
 for $N\rightarrow\infty$ that are uniform in a disk in the complex
plane \cite{Girko_et_al}. The radius $r$ of the disk is given by
\begin{equation}
r=N^{\frac{1}{2}}\sigma\label{eq:Gaussian_variance}\end{equation}
where $\sigma^{2}=\left\langle J_{ij}^{2}\right\rangle $ is the variance
of the matrix elements. 

Interestingly, we find that the radii of the eigenvalue distributions
of the above stability matrices (\ref{eq:IFmatrixelements}) well
agree with the radii obtained from Eq.~(\ref{eq:Gaussian_variance})
if $\left\langle J_{ij}^{2}\right\rangle $ is replaced by the variance
of the elements of the stability matrices shifted such that they also
exhibit a zero average eigenvalue. To directly compare the eigenvalues
of the stability matrices, which have average eigenvalue $\left[\lambda_{i}\right]:=\frac{1}{N}\sum_{i=1}^{N}\lambda_{i}=A_{0}$
to those of the Gaussian ensemble, we transform $A'_{ij}=A_{ij}-\delta_{ij}A_{0}$
shifting the average eigenvalues to $\left[\lambda'_{i}\right]=0$.
Here $\delta_{ij}$ denotes the Kronecker delta, $\delta_{ij}=1$
if $i=j$ and $\delta_{ij}=0$ if $i\neq j$. For the variance of
$A'$ we obtain\begin{eqnarray}
\sigma_{A'}^{2} & = & \left[A'_{ij}{}^{2}\right]-\left[A'_{ij}\right]^{2}\\
 & = & \frac{1}{N}\left(\sum_{\mysubstack{j=1}{j\neq i}}^{N}A_{ij}^{2}-\frac{(1-A_{0})^{2}}{N}\right).\label{eq:varianceA'}\end{eqnarray}
 For identical non-zero coupling strengths, the off-diagonal sum is
exactly equal to $\sum_{j=1,j\neq i}^{N}A_{ij}^{2}=(1-A_{0})^{2}/k$
such that, using (\ref{eq:Gaussian_variance}), we obtain the random
matrix theory prediction\begin{equation}
r_{\textrm{RMT}}=N^{\frac{1}{2}}\sigma_{A'}=(1-A_{0})\left(\frac{1}{k}-\frac{1}{N}\right)^{\frac{1}{2}}.\label{eq:rRMT}\end{equation}
for the radius $r$ of the disk of eigenvalues of the stability matrices
$A$ \cite{exact}. 

\begin{figure}
\begin{center}\includegraphics[%
  width=0.90\columnwidth]{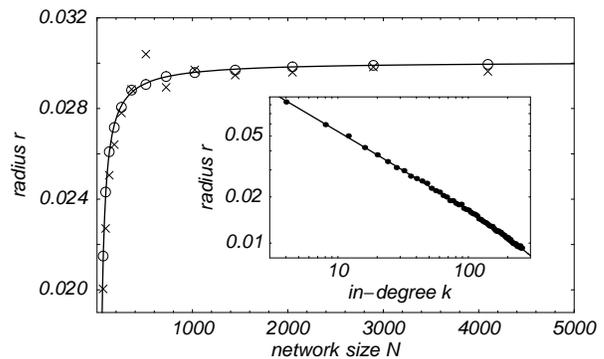}\end{center}

\caption{Scaling of the radius $r$ of the disk of non-trivial eigenvalues.
Main panel displays the radius $r$ as a function of network size
$N$ for fixed $k=32$. Symbols display and $r_{\textrm{Re}}$ ($\times$)
and $r_{\textrm{av}}$ ($\bigcirc$). Inset displays $r$ as a function
of $k$ for fixed $N=1024$. Dots display numerical data of $r_{\textrm{av}}$.
In the main panel and the inset, lines are the theoretical estimate
$r_{\textrm{RMT}}$ (Eq.~(\ref{eq:rRMT})) Other parameters as in
Fig.~\ref{Fig:EWSfreek8N}.\label{Fig:scalingk32N1024}}
\end{figure}

We verified this scaling law (\ref{eq:rRMT}) for various dynamical
parameters $A_{0}$ (determined by different $I$, $\eps$, and $\tau$),
network sizes $N$, and in-degrees $k$ and found excellent agreement
with numerically determined eigenvalue distributions, see, e.g., Fig.~\ref{Fig:EWSfreek8N}.
To quantify the accuracy of the prediction (\ref{eq:rRMT}) we numerically
estimated the radius of the distribution of the non-trivial eigenvalues
of $A$ for various $N$, $k$ as well as $A_{0}$. Results from two
different estimators are shown in Fig.~\ref{Fig:scalingk32N1024}.
The real part estimator $r_{\textrm{Re}}:=\frac{1}{2}\left(\max_{i\neq1}\MyRe(\lambda_{i})-\min_{i\neq1}\MyRe(\lambda_{i})\right)$
estimates the radius from the maximum spread of eigenvalues parallel
to the real axis. Typically, $r_{\textrm{Re}}$ should give an estimate
that is slightly inaccurate because it is based on two eigenvalues
only. This is circumvented by the average estimator $r_{\textrm{av}}:=\frac{3}{2}\frac{1}{N-1}\sum_{i=2}^{N}|\lambda_{i}-(A_{0}-(1-A_{0})N^{-1})|$
that estimates the radius $r$ of a circle from the average distance
$\left\langle d\right\rangle $ of eigenvalues from its center, because
$\left\langle d\right\rangle =\int_{0}^{2\pi}\int_{0}^{r}r'^{2}\rho(r')drd\varphi=\frac{2}{3}r$
if we assume a uniform density $\rho(r')$ according to (\ref{eq:uniform_EW_distribution}).
Here we take the center of the disk to be the average $\left\langle \lambda_{i}\right\rangle _{i\geq2}=A_{0}-(1-A_{0})N^{-1}+\mathcal{O}(N^{-2})$
of the non-trivial eigenvalues. Varying $k$ at fixed $N$ as well
as $N$ at fixed $k$ yields excellent agreement between the numerical
data and the theoretical predictions for sufficiently large $N$ and
$k$ (Fig.~\ref{Fig:scalingk32N1024}).%
 Varying the coupling strength $|\eps|$ and thus $A_{0}$ yields
equally good agreement (cf.~Fig.~\ref{Fig:tausyn}). 

The radius (\ref{eq:rRMT}) implies a prediction for the synchronization
time (see (\ref{eq:delta_tau})) \begin{equation}
\tau_{\textrm{syn}}=-1/\ln(A_{0}+r_{\textrm{RMT}})\,,\label{eq:tsyn}\end{equation}
 in terms of the (in modulus) largest non-trivial eigenvalue $\lambda_{\textrm{m}}$
. With increasing coupling strength $|\eps|$, the synchronization
time decreases. However, the speed of synchronization $\tau_{\textrm{syn}}^{-1}$
is bounded by a finite speed for arbitrary large $|\eps|$: Even if
$|\eps|\gg1$ and thus $A_{0}\ll1$, the largest non-trivial eigenvalue
asymptotically becomes $\lambda_{\textrm{m}}\approx k^{-1/2}$ for
sufficiently large $N$. Thus the shortest synchronization time \begin{equation}
\tau_{\textrm{syn }}^{|\eps|\rightarrow\infty}=\frac{2}{\ln k}\label{eq:taulimit}\end{equation}
 is limited by the network connectivity (cf.\  the asymptote in Fig.
\ref{Fig:tausyn}). This means that even for arbitrary strong interactions,
the speed of synchronization stays finite. Furthermore, at fixed $k$,
the synchronization time also cannot exceed a certain maximum, even
if the network size $N$ becomes extremely large (cf.\  Fig.~\ref{Fig:scalingk32N1024}).
This bound $\tau_{\textrm{syn}}^{N\rightarrow\infty}$ is determined
by the asymptotic radius $r_{\infty}:=\lim_{N\rightarrow\infty}r_{\textrm{RMT}}=(1-A_{0})k^{-1/2}.$
Moreover, because eigenvalues change continuously with a structural
perturbation to the system's dynamics, the existence of a gap $g:=1-(A_{0}+r_{\infty})>0$
indicates that no eigenvalue crosses the unit circle for sufficiently
small structural perturbations. Thus stable synchrony is not restricted
to the specific model considered here, but persists in systems obtained
by structural perturbations of the dynamics.

The above results show, that the distribution of eigenvalues of a
\textit{sparse} stability matrix with deterministic non-zero entries
at certain random positions is well described by the eigenvalue distribution
of the \textit{Gaussian} ensemble, which consists of fully occupied
matrices with purely random entries. This sparse-Gaussian coincidence
for asymmetric matrices is similar to that of symmetric random matrices
for large $k$: Gaussian symmetric matrices exhibit an eigenvalue
distribution $\rho_{\textrm{Gauss}}^{\textrm{s}}(\lambda)$, the Wigner
semicircle law \cite{Mehta}. Sparse symmetric matrices \cite{RodgersBray}
exhibit an eigenvalue distribution $\rho_{\textrm{sparse}}^{\textrm{s}}(\lambda)$
that is different from the semicircle law but approaches it in the
limit $k\rightarrow\infty$. For $k\gg1$ the distribution of eigenvalues
of sparse asymmetric random matrices $\rho_{\textrm{sparse}}^{\textrm{a}}$
appear to be well approximated by the eigenvalue distribution of Gaussian
asymmetric matrices, $\rho_{\textrm{sparse}}^{\textrm{a}}(\lambda)\approx\rho_{\textrm{Gauss}}^{\textrm{a}}(\lambda)$.
Our results indicate, that this is true even for moderate $k\approx10$. 

Further investigations of eigenvalue distributions for small-world
networks show that with decreasing randomness the speed of synchronization
decreases (the second largest non-trivial eigenvalue increases) such
that completely random networks synchronize faster than small-world
networks, at least asymptotically.

In conclusion, we have derived accurate analytical predictions for
the (asymptotic) speed of synchronization in asymmetric random networks
of oscillators in dependence of the dynamical parameters $\eps$,
$\tau$, $I$, as well as the network parameters $N$ and $k$. Even
the scaling with network size $N$, artificially introduced via the
variance (\ref{eq:varianceA'}) of finite matrices, is also accurately
reproduced (see e.g.~Fig.~\ref{Fig:scalingk32N1024}). As a particular
application, we explained the intriguing phenomenon, that the speed
of synchronization stays finite, even if the coupling strengths become
infinite. It turned out that the speed is restricted by the network
connectivity. In addition, at fixed parameters the synchronization
time does not increase above a finite threshold if the network becomes
very large. Furthermore, the existence of a gap between the non-trivial
eigenvalues and the unit circle indicates that stable synchrony is
a robust form of collective dynamics in a large class of systems.

Random matrix theory has previously been applied to various physical
systems that exhibit certain symmetries -- such as time-reversal symmetry
-- but an otherwise unknown structure. For instance, correlations
of energy levels in nuclear physics and quantum mechanical properties
of classically chaotic systems have been successfully predicted (see
Ref.~\cite{Forrester} for a recent review). Our results demonstrate
that random matrix theory also is an appropriate tool for analyzing
synchronization in random networks of dynamical units. Possible lines
for future applications may include synchronization phenomena of pulse-
and phase-coupled units as well as of chaotic dynamical systems (cf.~also
\cite{Gade}). The limits of synchronization speed predicted in this
Letter, are expected to occur in those systems, too. More generally,
other equilibration processes and the dynamics in more structured
topologies such as small-world networks may be analytically investigated
using statistical spectral properties of the respective operators,
too.

We thank T. Kottos, P. Müller, H. Sompolinsky, M. Weigt, and A. Zippelius
for useful discussions.


\begin{thebibliography}{10}
\bibitem{Albert_et_al}S. Bornholdt and H. G. Schuster (eds.), \textit{Handbook of Graphs
and Networks}, (Wiley-VCH, Weinheim, 2002); R. Albert and A.-L. Barab\'{a}si,
Rev. Mod. Phys. \textbf{74}, 47 (2002); S. N. Dorogovtsev and J. F.
F. Mendes, Adv. Phys. \textbf{51}, 1079 (2002); M. E. J. Newman, SIAM
Review \textbf{\noun{45}}, 167 (2003). 
\bibitem{Strogatz}S. H. Strogatz, Nature \textbf{410}, 268 (2001). 
\bibitem{Nadis}S. Nadis, Nature \textbf{421}, 780 (2003).
\bibitem{Pikovsky}A. Pikovsly, M. Rosenblum, and J. Kurths, \textit{Synchronization:
A Universal Concept in Nonlinear Science} (Cambridge University Press,
Cambridge, England, 2003).
\bibitem{Wiesenfeld}K. Wiesenfeld, P. Colet, and S. H. Strogatz, Phys. Rev. Lett. \textbf{76},
404 (1996). 
\bibitem{syn}C. M. Gray, P. König, A. K. Engel, and W. Singer, Nature \textbf{338},
334 (1989); R. Eckhorn \textit{et al.} Biol. Cybern. \textbf{60} 121
(1988); W. Singer, Neuron \textbf{24}, 49 and 111 (1999).
\bibitem{Winfree_et_al}A. T. Winfree, \emph{The Geometry of Biological Time} (Springer, New
York, 1980); Y. Kuramoto, \emph{Chemical Oscillations, Waves, and
Turbulence} (Springer, Berlin, 1984); for a recent review on the Kuramoto
model see S. H. Strogatz, Physica D \textbf{143}, 1 (2000).
\bibitem{Abbott_et_al}A. V. M. Herz and J. J. Hopfield, Phys. Rev. Lett. \textbf{75}, 1222
(1995);  \'{A}. Corral, C. J. P\'{e}rez, A. D\'{i}az-Guilera,
and A. Arenas, Phys. Rev. Lett. \textbf{75}, 3697 (1995); W. Gerstner,
J. D. Cowan, and J. L. van Hemmen, Neural Comp. \textbf{8}, 1653 (1996);
P. C. Bressloff, S. Coombes, and B. de Souza, Phys. Rev. Lett. \textbf{79},
2791 (1997); C. van Vreeswijk, Phys. Rev. Lett. \textbf{84}, 5110
(2000); D. Hansel and G. Mato, Phys. Rev. Lett. \textbf{86}, 4175
(2001).
\bibitem{Baharona}M. Barahona and L. M. Pecora, Phys. Rev. Lett. \textbf{89}, 054101
(2002).
\bibitem{TimmePRL}M. Timme, F. Wolf, and T. Geisel, Phys. Rev. Lett. \textbf{89}, 258701
(2002).
\bibitem{Earl}M. G. Earl and S. H. Strogatz, Phys. Rev. E \textbf{67}, 036204 (2003).
\bibitem{Mirollo_et_al}R. E. Mirollo and S. H. Strogatz, SIAM J. Appl. Math. \textbf{50},
1645 (1990); U. Ernst, K. Pawelzik, and T. Geisel, Phys. Rev. Lett.
\textbf{74}, 1570 (1995); M. Timme, F. Wolf, and T. Geisel, Chaos
\textbf{13}, 377 (2003).
\bibitem{uniqueness}In general, the stability matrix depends on the order of incoming
signals. In this Letter, however, we focus on a subclass of models,
for which this complication does not arise because the stability matrix
becomes unique \cite{TimmeDoc}.
\bibitem{TimmeDoc}M. Timme, \emph{Collective Dynamics in Networks of Pulse-Coupled Oscillators},
Doctoral Thesis, University of Göttingen, 2002.
\bibitem{Stoer}J. Stoer and R. Burlisch, \textit{Introduction to Numerical Analysis}
(Springer, Berlin, 1992).
\bibitem{Girko_et_al}V. L. Girko, Theory Probab. Appl. \textbf{29}, 694 (1985). H. J. Sommers,
A. Crisanti, H. Sompolinsky, and Y. Stein, Phys. Rev. Lett. \textbf{60},
1895 (1988).
\bibitem{exact}Here $k=k_{i}$ for all $i$ such that the estimate for the radius
(\ref{eq:rRMT}) is exact. If the random network is constructed by
choosing every connection independently with probability $p$, it
becomes an approximation if $k^{-1}$ is replaced by $\left[k_{i}^{-1}\right]_{i}$.
\bibitem{RodgersBray}G. J. Rodgers and A. J. Bray, Phys. Rev. B \textbf{37} 3557 (1988).
\bibitem{Forrester}P. J. Forrester, N. C. Snaith, and J. J. M. Verbaarschot, J. Phys.
A \textbf{36}, R1 (2003).
\bibitem{Mehta}M. L. Mehta, \emph{Random Matrices} (Academic Press, New York, 1991).
\bibitem{Gade}P. M. Gade, Phys. Rev. E \textbf{54}, 64 (1996).\end{thebibliography}
\end{document}